\RequirePackage{fix-cm}
\documentclass[smallextended]{svjour3}       
\smartqed 
\usepackage{pdflscape}
\usepackage{graphicx}
\usepackage{lscape}
\usepackage{csquotes}
\usepackage{natbib}
\usepackage[a4paper, total={6.10in,9.25in}]{geometry} 
\begin{document}
	
	\title{Recommendation Systems for Tourism Based on Social Networks: A Survey 
	}
	
	
	\author{Alan Menk \and Laura Sebastia \and Rebeca Ferreira}
	\institute{Universitat Politecnica de Valencia\at
		Valencia - Spain
		Cami de Vera s/n \\
		Tel.: +34 96 387 70 00\\
		\email{amenkdossantos@dsic.upv.es}\\
		\email{lstarin@dsic.upv.es}\\				
		\email{redelfer@posgrado.upv.es}	
	}
	\date{Received: date / Accepted: date}
	\maketitle

\begin{abstract}
Nowadays, recommender systems are present in many daily activities such as online shopping, browsing social networks, etc. Given the rising demand for reinvigoration of the tourist industry through information technology, recommenders have been included into tourism websites such as Expedia, Booking or Tripadvisor, among others. Furthermore, the amount of scientific papers related to recommender systems for tourism is on solid and continuous growth since 2004. Much of this growth is due to social networks that, besides to offer researchers the possibility of using a great mass of available and constantly updated data, they also enable the recommendation systems to become more personalised, effective and natural. This paper reviews and analyses many research publications focusing on tourism recommender systems that use social networks in their projects. We detail their main characteristics, like which social networks are exploited, which data is extracted, the applied recommendation techniques, the methods of evaluation, etc. Through a comprehensive literature review, we aim to collaborate with the future recommender systems, by giving some clear classifications and descriptions of the current tourism recommender systems.
\end{abstract}

\section{Introduction}
The expansion of the use of Social Networks (SNs) has become clear when looking at the increase in the number of users and the data volume in these networks. In 2016, 79\% of U.S. adults that accessed the internet used the SN Facebook \citep{greenwood2016}. As a result, the amount of data available for several purposes (marketing, investigation, analysis, etc.) is vast. The Digital Universe study from the International Data Corporation (IDC) found that, back in 2012, about 68\% of data was created and consumed by users watching digital TV, interacting with the social networks, sending images and videos taken with the phone camera between devices and around the internet. Their prediction is that, in 15 years, the world's data will grow by a factor of 300 \citep{iview2012}.
	
The continuous changes in the consumption behaviour of the individuals in the internet or in the way we communicate with family and friends are evident. Companies of products and services spend increasing amounts of money in marketing and advertising with the focus on the web, in detriment of traditional physical means of communication, like newspapers and flyers. Therefore, Recommendation Systems (RSs) are becoming more present in many websites and applications, such as SNs (e.g. Facebook), e-commerces (e.g. Amazon), so that we are recommended where to travel to (e.g. Expedia), what music to listen (e.g. Spotify), what films to watch (e.g. Netflix), what to eat (e.g. Ifood) or even who date with (e.g. Tinder) \citep{linden2003, kabiljo2015, das2007}.
	
The main input that allow RSs to work is data about user tastes and preferences. How to access these data was one of the main bottlenecks of RSs some time ago. Now, a relevant amount of the population is connected to SNs and, therefore, data about users (usually highly intimate and personal) can be accessed more easily. With the use of these data, an RS can \enquote{learn} about what a particular user likes, but more than this, it can analyse intrinsic and personal information such as his psychological profile, context issues, a profile of his circle of friends, etc. The current RSs are becoming more accurate, and they can be integrated with target platforms to obtain data in an implicit way, so it is possible to replace data acquisition through lengthy forms, annoying questions, etc. Consequently, the cold-start problem, a well-known and discussed issue in RSs, can be mitigated when users connect to a new platform by using a pre-existent account in another platform, thus enabling it to have access to the data available in the first one. In brief, what we highlight here is the importance that SNs can have in RSs.
	
In parallel, the area of tourism is an important source of income of countries and regions. Nowadays 10\% of GDP corresponds to a direct, indirect or induced effect on tourism, creating 1 out of 11 jobs in a country. In 1950, the number of international tourists in the world was 25 million whereas, in 2015, it jumped to 1186 million and the predictions for 2030 are 1800 million, an annual growth of 3.3\% \citep{unwto2016}. Tourism is the first or second source of income in the economy in 20 of the 48 least developed countries. In European countries like Spain, the activity reached numbers like 10.9\% of GDP in 2014, that is 12.7\% of the jobs in this country was possible thanks to tourism. This favourable scenario, where users and data in SNs are massive, made many important projects arose \citep{ting2016} in order to build more efficient and customised systems in the tourism sector.
	
In this work, we present a review of existing tourism RSs that use data from SNs and discuss some research directions following the same ideas. The motivations of this work are diverse:
	
\begin{itemize}
\item SNs have been a {\em popular research} area, not only regarding data and web mining but also with respect to SN analysis. Many of these works are devoted to develop new techniques and algorithms or to improve traditional mining techniques for SN analysis, decision support and RSs \citep{ting2016}. In fact, very profitable knowledge for RSs can be obtained from SNs due to the rising amount of data available online, and from the analysis of social relations existing in the web, that mainly reflect the behaviour of the real world \citep{amichai2010}, providing an opportunity to study them through computational algorithms.
		
\item SNs can help to {\em improve the prediction accuracy} of RSs in two ways. First of all, the quality of the available data can offer detailed information about the users, including their preferences, tastes, and social or geographical context. The second point is related to the possibility of predicting the user personality from data available in SNs, which could be specially valuable for particular market niches and recommendation systems. Moreover, in both cases, data are obtained implicitly, thus avoiding the use of long forms or tests.
		
\item By reviewing the existing studies, it is possible to get to know the {\em approaches and methods} used by the researchers when introducing/combining data from SNs into RSs, thus gathering the best practices and using them as a starting point to keep improving the items to recommend, consequently increasing the satisfaction of the individuals.

\item A {\em classification} of tourism RSs that use SNs can help developers and researchers gain a quick understanding of which kind of data can be retrieved from SNs and are the most used to generate recommendations.
\end{itemize}
	
	
	

	
This paper is organised as follows. First, the basic techniques used in RSs are described; then, we give a brief overview on SNs and their types (Section 2). Section 3 describes the methodology used for selecting the papers reviewed in this work. In Section 4, we present an analysis of the selected papers, with a general classification of them and a temporal evolution of SNs, comparing with the amount of papers related to them. We also try to answer some questions using the analysed projects such as what data are extracted from SNs, which the main used recommendation techniques are, which type of recommendations are generated and how they are presented to the user and which evaluation methods are used. Finally, we present the discussions, conclusions, research challenges and future prospects (Section 5).

\section{Background}
	
\subsection{Techniques for recommender systems}
RSs are software tools and techniques that provide suggestions of items that are most likely of interest to a particular user \citep{kantor2011}. Studies about recommendations, suggestions or content filtering for the tourism sector are not that new. In 1986, \citep{michie1986} proposed that travellers construct their preferences for alternative destinations from their awareness and effectiveness; in 1989, \citep{woodside1989} proposed a path model of direct and indirect relationships leading to destination choice. In the mid-1990's, \citep{woodside1990} presented a framework of routes selection in Prince Edward Island region (Canada). The authors developed propositions suitable for empirical testing by using eight leisure traveller choice subsystems: destinations, accommodations, activities, visiting attractions, travel modes, eating options, destination areas, and routes. However, it is worth mentioning that they reported the data collection as their biggest limitation, which was made entirely manually, but also the amount of available personal data about travellers, actually hardly null. From this century on, with continuously increasing rates of new users on the web, surrounded by the beginning of the mobile age, the problem of lack of data faced in the 90's in the projects about the recommendation in the tourism sector is not a problem anymore. This section describes a summary of the main techniques used in RSs.
	
\textit{Content-Based (CB)}: Essentially, a CB RS learns to recommend items that are similar to those the user has liked in the past. The similarity of items is calculated based on the features associated to the compared items. The main advantage of this technique is the \enquote{user independence}, given that it depends only on the user's own data; in other words, it identifies the common characteristics of items that have received a favourable rating from a user \textit{u}, and then it recommends to \textit{u} new items that share those characteristics \citep{pazzani1997, billsus2000, balabanovic1997}. For example, when a user rated (positively) a point of interest (POI), the system can recommend similar POIs by calculating how similar these two POIs are according to their features.
	
\textit{Collaborative Filtering (CF)}: It is the process of filtering or evaluating items using the opinions of other people \citep{schafer2007}. These opinions can be obtained explicitly from users through form responding, or by using some implicit measures, such as records of previous purchasing. That is, CF is an algorithm for matching people with similar interests for the purpose of making recommendations \citep{ricci2011}. For instance, a system may recommend a customer who travelled to Paris and Barcelona, to travel to Rome, because other users that travelled to Paris and/or Barcelona, travelled to Rome as well. Two types of CF algorithms can be found: (1) memory-based CF, where user rating data is used to compute the similarity between users or items and (2) model-based CF, where models are developed using different data mining and machine learning algorithms to predict users' rating of unrated items.
	
\textit{Knowledge-Based (KB)}: This technique works by recommending items based on specific domain knowledge about how certain item features meet users' needs and preferences and, ultimately, how the item is useful for the user \citep{ricci2011}. In other words, it generates recommendations to the user based on the knowledge about his needs towards a particular item. These recommendations are performed under measures of utility, derived from the knowledge of the relationship between a specific user and item. For instance, a KB tourism RS will generate recommendations not only based on the past travel experience of the user, but also based on what are the characteristics of the places/cities visited and the places available to recommend, that is, a KB RS exploits knowledge to map a user to the products he likes. They can use a wide range of techniques and, at the same time, they require a big effort in terms of knowledge extraction, representation and system design.
	
\textit{Demographic Filtering (DF)}: Essentially, this algorithm recommends items based on the demographic profile of the user (Bobadilla et al, 2013). In other words, this technique provides different recommendations for different demographic niches, combining the ratings of users in these niches \citep{ricci2011, pazzani1999, wang2012}.
	
Finally, we also find hybrid RSs which are based on the combination of the above mentioned techniques \citep{ricci2011} (or some others, because this is not an exhaustive list). A hybrid RS combines techniques \enquote{X} and \enquote{Y} trying to enhance the advantages of \enquote{X} to mitigate the disadvantages of \enquote{Y} (and vice versa).
	
Nowadays, there is a great variety of techniques, models, algorithms, etc. that are used in different RSs. For example, the context-aware RSs, that characterise the situation of an entity (person, place or object) that is considered relevant to the interaction between a user and an application, including the user and the application itself \citep{abowd1999}. For instance, in a tourism RS, the context referring to the season in which a person is going to travel is important because recommendations of destinations in winter should be very different from those provided in summer \citep{balabanovic1997}.

\subsection{Social Networks}
SNs are means of electronic communication through which users create online communities to share information, ideas, personal messages, and other content (as videos) \citep{merriam-webster2015}. To define a web page as a SN, it must cover three essential characteristics: to offer services that allow individuals to construct a public or semi-public profile within a bounded system, to articulate a list of other users with whom they share a connection, and to offer the opportunity of viewing and traversing their list of connections and those made by others within the system \citep{ellison2007}.
	
\begin{figure}[h]
	\centering
	\includegraphics[scale=0.4] {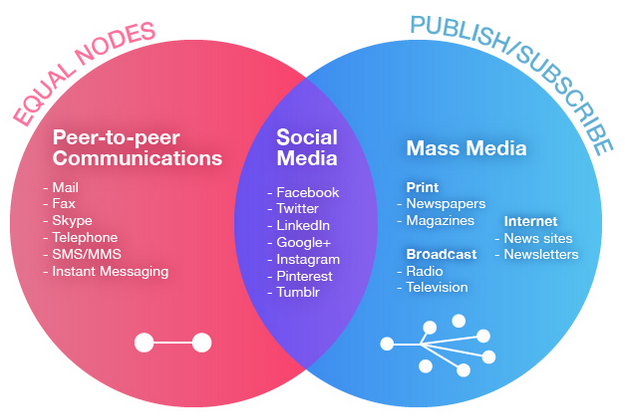}
	\caption{Relation between peer-to-peer communication and mass network generating the SN by \citep{veerasamy2013}}
	\label{socialnetwoks}
\end{figure}

There are further definitions, such as from \citep{kaplan2010}, who define it as a group of internet-based applications that build on the ideological and technological foundations of Web 2.0, and allow the creation and exchange of user-generated content. Also, for the authors, the SNs are applications that enable users to connect by creating personal information profiles, inviting friends and colleagues to have access to their profiles, and sending e-mails and instant messages between each other. In brief, a SN is a structure composed of people or organisations that share values and common goals. Figure \ref{socialnetwoks} represents, on the one hand, the individual means of communication \textit{(1 to 1)}, like, for example, phones and internet telephony service providers (such as Skype); and, on the other hand, the mass media \textit{(1 to n)} like TV, radio, printed or online newspapers and magazines. Finally, if these two scenarios are combined, SNs \textit{(n to n)} emerge, as we know them today.
	
Since their creation, SNs have been producing an astonishing amount of data, as previously mentioned. Such growth is not merely regarding the available content, but also the growing use of internet and consequently of SNs. For instance, in the middle of 2015, Facebook reached a 1.5 billion of users who have used it at least once in a month; this means that one in seven people in the world connected to Facebook in 2015.

Nowadays, even with increasingly restrictive policies, it is possible to obtain not only standard data widely used in traditional forms (i.e. name, age, gender, marital status) but also information extremely \enquote{intimate} about users, as personal preferences, likes, past trips or even where the person wants to travel to. With such valuable information available in SNs, we understand they can enrich and improve the predictions of RSs in the tourism sector.

\section{Methodology}
	
	
As explained above, our aim in this survey is to analyse existing works on tourism RSs that use data from SNs. Our search of scientific papers was performed by means of a filtering process in several databases such as: ACM Digital Library, IEEE Xplore, dblp, Emerald, Springer Link, Science Direct, Web of Science, Scopus, Dialnet plus, among other open source databases like DOAJ. Only articles and e-books were selected as document types, and we only selected RSs, searched as \enquote{recommend*}, oriented/aimed to the tourism sector (\enquote{touris*}) that used some type of data from SN in their model (\enquote{social network*}).
	
\begin{figure}[h]
	\centering
	\includegraphics[scale=0.65] {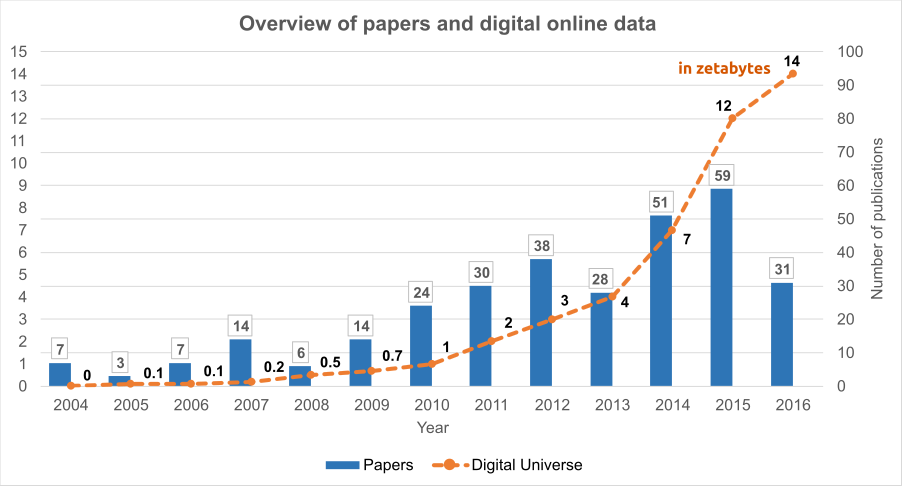}
	\caption{The growth of online data \citep{gantz2012} and tourism RSs research based on SNs.}
	\label{2020}
\end{figure}
	
Figure \ref{2020} shows a summary of the result of this search. Here, we can observe the relation between the number of scientific papers found in our search, as well as the amount of online available data. The number of publications was insignificant until the year 2004, so they are not depicted in Figure \ref{2020}. However, with the expansion of the use of SNs in 2004, data (text, video, audio and other files) started to increase, albeit the amount of related papers oscillated between 3 and 14 in the subsequent 5 years. From 2009 onwards, the growth in the research represented in Figure \ref{2020} is clear, which can be associated to the growth in the volume of data available online in zetabytes (thanks to the inclusion of new devices such as tablets and the increasing of the number of smartphones), in addition to the launching/release of APIs for the main SNs. In 2009 and 2010, the scientific papers found rised from 14 to 24 and kept growing until reaching the peak of 59 papers in 2015. In the following year, though, only 31 projects were found, which could be related to the limitation on the access to the main SNs' data through their APIs. An example of this data limitation is Facebook, that limited the access to users' data in 2015 \citep{seetharaman2015}. Other SNs such as Twitter and Linkedin are also putting restrictions in the data accessible through their APIs. In parallel, we should not be disregarded that we can see an increase in the volume of data (text, video, audio and others files) from 2004 to 2016, when we reached 14 zettabytes of data generated on the internet \citep{kanellos2016}.
	
From the 312 papers that fulfilled our search parameters, we selected 31 papers to be deeply analysed, following three criteria: those which used the most known SNs (based on the number of users); those focused on the development of a practical application, that is, real RSs; and those with a relevant number of citations. Tables \ref{generaltable1} and \ref{generaltable2} show the list of selected papers descendingly ordered by the year of publication.
	
	
Once defined the target work to be studied, we classified these papers by the different aspects that we wanted to analyse, which are shown in the columns of Tables \ref{generaltable1} and \ref{generaltable2}. Specifically, these aspects are:

\begin{enumerate}
	\item SNs, that is, from which SN data is extracted.
	\item Other data sources, that is, additional data sources used in these papers (if any).
	\item Extracted items, that is, which type of information is extracted from SNs.
	\item Recommendation technique, which indicates the several recommendation techniques used in each system.
	\item Evaluation, indicating whether the evaluation of the system was performed using synthetic data or real users.
	\item Recommendation system properties, describing which desirable properties are pursued in each system, such as accuracy, serendipity, etc.
	\item Output, regarding whether the RS shows a list of POI recommendations, a route or a guide.
	\item Interface used by the user to interact with the RS.
\end{enumerate}
	
We also discuss the relevance of each aspect and the positives of the main systems, which will be detailed in the next sections.
	
\begin{table}[h]
	\centering
	\begin{tabular}{c}
		\includegraphics[scale=0.29] {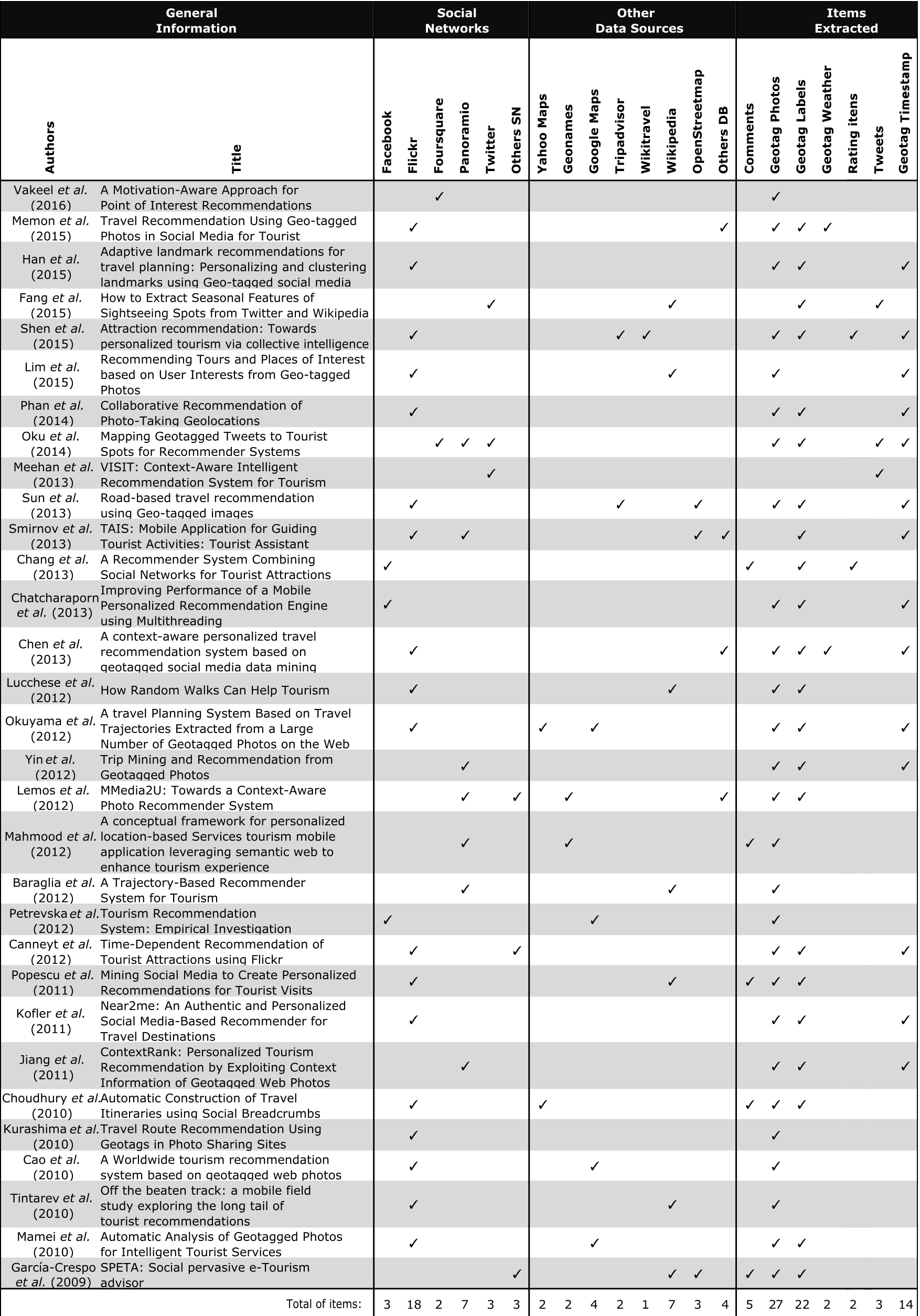}
	\end{tabular}
	\caption{Overview of classified projects and their characteristics sorted by date of publication, showing the SNs used and items extracted.}
	\label{generaltable1}
\end{table}

\section{Analysis of Tourism Recommender Systems using Social Networks}

\subsection{Which social networks and additional data sources are used?}
This section gives a brief review of the main SNs employed in tourism RSs in the last years, which are summarised in the second column of Table \ref{generaltable1}. We found projects that work with widely used SNs such as Facebook and Twitter, and others focused on a more specialised audience such as Flickr\footnote{{\tt www.flickr.com}}, that allows the user to store, search, sell and share photos or videos; Foursquare\footnote{{\tt www.foursquare.com}}, a local search-and-discovery service mobile app which provides personalised recommendations of places to go to near a user's current location based on users' previous browsing history, purchases, or check-in history and Traveleye\footnote{{\tt www.traveleye.com}}, focused on trips organisation, that allows users to write posts with travel experiences, to follow other travellers' journeys, to share travels with friends, to search tourist attractions and travel guides, etc. We also have found works that used no longer available SNs, such as Picasa \citep{lemos2012} and Panoramio, which was a was a geo-located tagging, photo sharing mashup, acquired by Google in 2007.

In relation to the analysed projects, we can observe that Flickr was the most used, with 58\% of the projects \citep{shen2015, han2015, memon2015}, among others; then, Panoramio with 23\% \citep{smirnov2013, jiang2011}. The advantage of these SNs is that they enable the collection of \enquote{Coordinates}, some \enquote{Geotag Labels} and even data about the person who took the photo, providing researchers with interesting data for RSs.

Facebook and Twitter (1st and 5th most used in the world \citep{social2016}) are used only in the 10\% of the analysed works. This low rate can be explained with the fact that both are generalist SNs and, therefore, data related to tourism is more difficult to obtain. Several works \citep{chang2013, chatcharaporn2013, petrevska2012} have used Facebook to obtain numbers of likes, groups, friends, comments or geotags of check-ins.

With regard to Twitter, the core data are the user tweets, retrieved with different goals. For instance, \citep{fang2015} considered the concept of sightseeing spots for different seasons, thus generating seasonal feature vectors for each sightseeing spot, which could support context-aware recommendation of tourist spots depending on the time of the year. Tweets also can be used to characterise the tourist spots \citep{oku2014}, or be combined with sentiment analysis to determine the current \enquote{mood} of each tourist \citep{meehan2013}. \citep{canneyt2011} opted to work with Twitter and Traveleye in their project. The first was employed for inferring the sentiment analysis merged with context-aware (location, weather and time) data, while Traveleye was used to extract the moment when the user visited a given city.
	
\begin{figure}[h]
	\centering
	\includegraphics[scale=0.67] {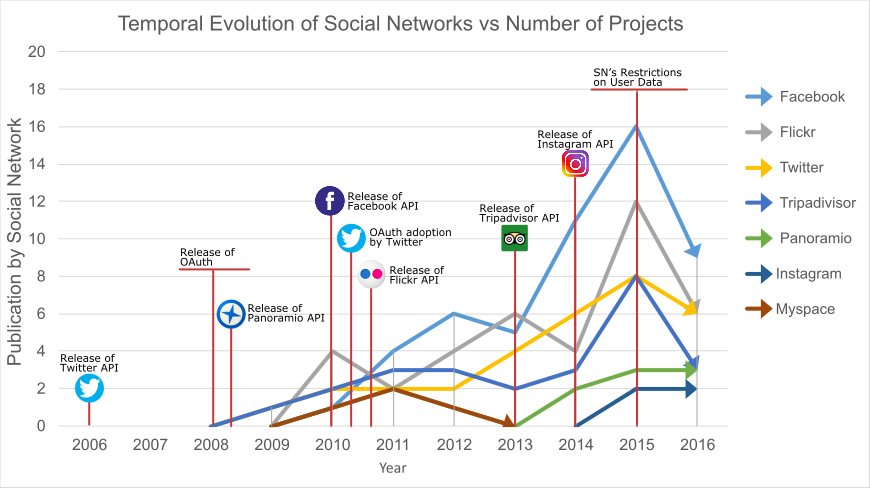}
	\caption{Temporal evolution depicting the papers of fig. \ref{2020} classified by SN and the year of appearance of their APIs and protocols.}
	\label{temporal}
\end{figure}

Figure \ref{temporal} shows the temporal evolution of some of the SNs that stand out in the recommendation projects oriented to the tourism sector since 2006, relating their appearance with the number of papers found in our search. In the middle of 2006, Twitter released an Application Programming Interface (API) for easing the access to data. Nowadays, all the major SNs have their own APIs, which allow to obtain data in an organised and automated way, by means of function calls. Combined with OAuth\footnote{An open standard for access delegation, which allows an end user's account information to be used by third-party services, such as Facebook, without exposing the user's password.}, released in 2008, APIs enable wider approaches of user integration, besides to add value to the user, the developer and the application. We observe that a number of projects, independent from the SN used, started to appear in 2008 and kept growing until 2015. Specifically, Facebook was used in 16 papers in 2015, followed by Flickr and Twitter, with 12 and 8 papers, respectively. One of the components that boosted such growth could be the ripeness of the available technologies with the definition of new standards, protocols and the documentation for their platforms. This opened an opportunity, even for non-IT researchers, to have access to data, integrate systems and develop new tools in a quick and simplified way. The reduction in the publications registered in 2016 seems to be related to the limitation on the access to users' data imposed by the main SNs, as explained previously. In summary, Figure \ref{temporal} shows the relation between the ease to access data from SNs after the standardization of access and authentication (APIs, OAuth, etc.) and the volume of published papers that use these SNs.

On the other hand, Table \ref{generaltable1} shows, in the third column, other data sources used in the analysed papers, as a complementary source. Most of the analysed projects used these additional data sources for showing the recommendations on a map. For example, \citep{okuyama2013, petrevska2012, sun2013} used Yahoo Maps, Google Maps and OpenstreetMaps, respectively. Others, such as \citep{lim2015} used Wikipedia to extract the list of POIs, latitude/longitude coordinates, and interest categories. \citep{lucchese2012} considered that the advantage of using Wikipedia is twofold. They used it, in one hand, to identify a large number of POIs in every city (even the less popular ones) and, on the other hand, to provide additional structured information about the POI (e.g. a subdivision of categories). \citep{sun2013} have chosen to use TripAdvisor to obtain a dictionary of landmarks. In this same scenario, \citep{shen2015} have used Tripadvisor to retrieve user comments about candidate attractions, besides the user rating about each attraction. \citep{smirnov2013}, in addition to Wikipedia and Panoramio, have also used another data source, the Wikivoyage, to obtain detailed information about the attraction.

\subsection{What data are extracted from social networks?}
Recommender Systems mainly need two types of information: information about user tastes and preferences and information about the items to recommend. In our analysis, we have noted that SNs are used for retrieving both. Regarding items, SNs can be used for discovering new items or for adding additional characteristics to existing items in the RS database.
	
In the last column of Table \ref{generaltable1}, we can see that, regardless the type of SN used in the reviewed projects, the collected data are quite similar: 87\% of them use an SN for obtaining \enquote{Geotag Photos}, that is, labels that contain the geographical identification metadata, such as latitude and longitude coordinates, though they can also include altitude, bearing and distance, accuracy data, like \citep{baraglia2012, memon2015}, among others; 71\% extract \enquote{Geotag Labels}, which are labels indicating the name of the city, country, address or labels that describe the photo, fundamental in projects such as \citep{memon2015, han2015}; and, finally, the \enquote{Geotag Timestamp}, which indicates when a photo (for example) was taken, is used in 45\% of the projects \citep{okuyama2013, yin2012}.
	
Less used, but also important, is textual information such as \enquote{Comments} and \enquote{Tweets}, which are used to extract keywords/labels commonly exploited in projects of text mining and sentiment analysis. \enquote{Comments} were used in 16\% of projects, like \citep{kofler2011, jiang2011} or \citep{chang2013}, which extracted items shared by the user in Facebook along with likes, comments and ratings. On the other hand, \citep{fang2015, meehan2013, oku2014} worked with tweets in their projects.

Only 6\% of papers used a \enquote{Geotag Weather}, tags that contain weather information for a particular location, which helps the development of context-aware systems \citep{lim2015, oku2014, baraglia2012}; the same figure has the \enquote{Rating Items} \citep{shen2015, chang2013}, that means, the extraction of ratings such as online evaluations made by the users, which indicates their level of satisfaction (e.g. stars, ranking, likes) 	regarding restaurants, hotels, cities, POIs, routes, etc.

Table \ref{generaltable1} shows that many projects combine several of these data. For instance, \citep{shen2015} collected heterogeneous data source using Flickr, Tripadvisor and Wikitravel: from Flickr, photos with metadata (time, location, attraction, and User ID); from Tripadvisor, user comments about the candidate attractions and user rating about each attraction; and from Wikitravel, official travelogues. Such heterogeneity reflects in the system a performance gain in terms of effectiveness as well as efficiency; an example is the \enquote{coordinates} of POIs, which is combined with \enquote{comments} and \enquote{rating items} from TripAdvisor, with the aim of learning from the experience of tourists who already visited the POI. In this case, collective intelligence is first gathered from a large amount of user-generated content in social media. Also, different aspects of knowledge can be mined from collective intelligence for denoising data and structuring heterogeneous information. In \citep{oku2014}, three different data sources were used: Foursquare, to obtain POI names, coordinates and category; Twitter, for the date, hour and coordinates of the visit; and Panoramio to obtain a POI photo with the title, coordinates and owner. In other words, three types of datasets were used: tourist spots, geotagged tweets, and geotagged photographs to generate a method for mapping geotagged tweets to tourist spots on the basis of the substantial activity regions of the spots and also for extracting temporal features and phrasal features based on the mapped tweets, with a positive level of effectiveness according to experiments developed.
	
The second target when obtaining data from SN is focused on the discovery of behavioural patterns, preferences, and personal characteristics of users. In this case, it is valuable the extraction of user profiles, friends, and comments, which are the three key components of SNs \citep{danah2007}. For example, \citep{garcia-crespo2009} recommend attractions that are likely to fit the current user expectations by exploiting the information exposed by user preferences; here, they based on the current user profile of the SN OpenSocial\footnote{http://code.google.com:80/apis/opensocial/}, which determines the common characteristics of the previously visited places and the user behaviour. Within this project, several elements were extracted from the SN: (1) Coordinates, which means knowing the user's location, allowing the offering of a set of places, but also the detection of contacts or friends in the surrounding areas; (2) Time and weather, to recommend indoor locations when the weather does not give any other possibility, also taking into consideration timetable restrictions of attractions; (3) The users' profile through the explicit interaction of the user, determining what their interests are, what kind of places they prefer to visit, and the ratings given to attractions; but also through the implicit data retrieval, collecting information regarding favourite painters, writers, or music preferences, for instance.
		
In this same line, the VISIT project \citep{meehan2013} used five types of contextual data, which are location, time, weather, social media sentiment and personalisation. The location is extracted from three main location sensing techniques used outdoors: GPS, GSM and WiFi; time, calculated from the amount of time that a user stays at each attraction; weather, extracted from the WorldWeatherOnline; social media sentiment, performed on Twitter messages (tweets) in real time to determine to current “mood” of each tourist attraction; and personalisation, by using the user profile data to describe a person in terms of age, gender, relationship status and the number of children, which can be used as a starting point for the application when first launched with no previous history.
	
As we can see, data extraction can be performed over a unique or multiple SNs, and in each case, one or more pieces of information about items that can be extracted. In addition, generally the use of data from SNs have some interesting advantages, such as the fact of counting on real data, the chance to later make tests with the users and also the availability of well-defined APIs provided by the most important SNs, thus making the development of their projects easier. However, we observe that regardless the SN, researchers face the same problem: irrelevant or false data, not only in case of users that insert or dismiss such information, but also with respect to those responsible for the development of SNs, who sometimes do not specify well the categories or standards neither establish required fields.
	
\begin{table}[h]
	\centering
	\begin{tabular}{c}
		\includegraphics[scale=0.29] {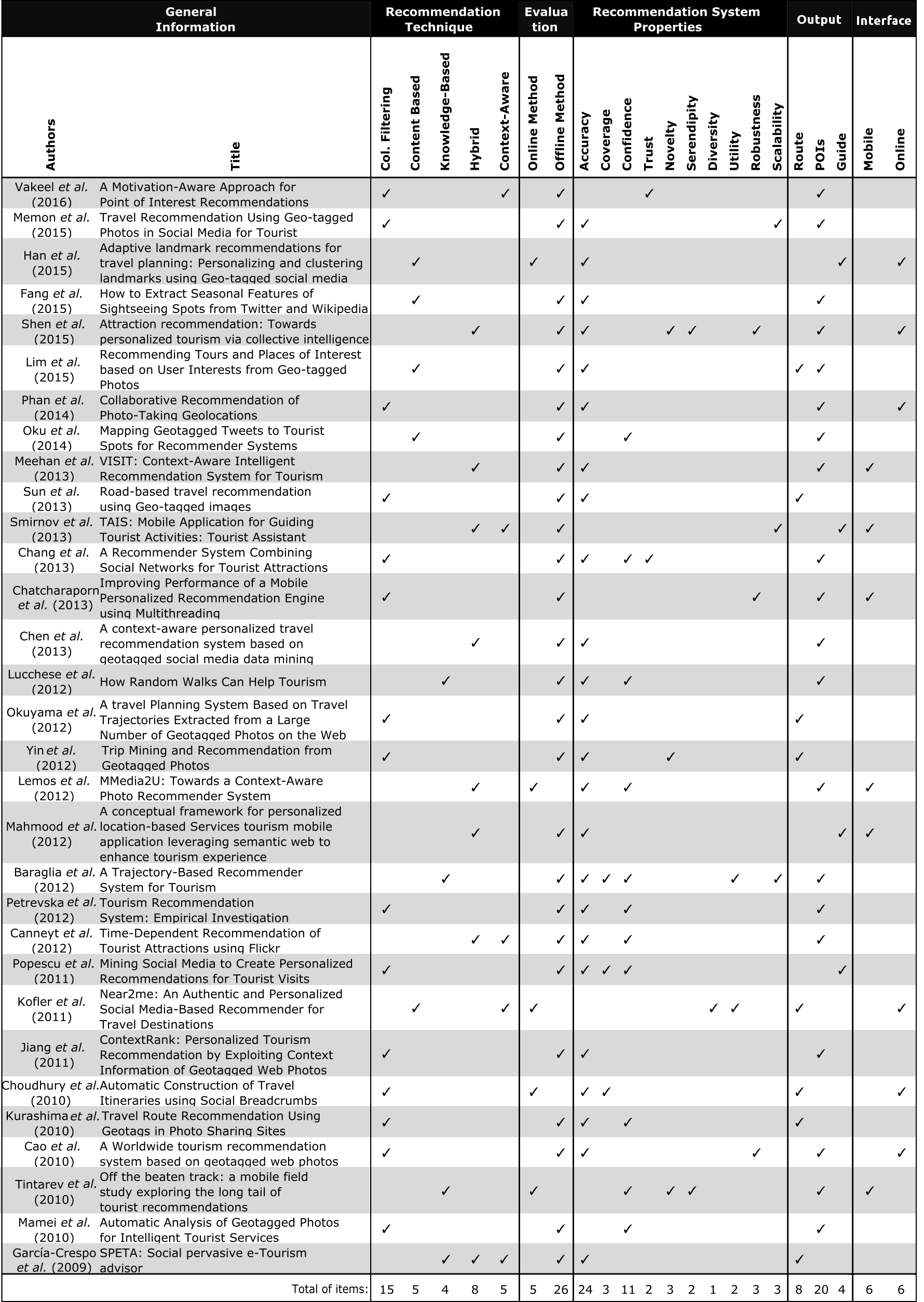}
	\end{tabular}
	\caption{Overview of classified projects and their characteristics sorted by date of publication, showing the evaluation, technique used, output, architecture and interface.}
	\label{generaltable2}
\end{table}

\subsection{What recommendation techniques are used?}
With respect to the recommendation techniques used in the papers that we have analysed, we distinguish between those using the more traditional techniques, such as content-based, collaborative filtering and knowledge-based techniques and those combining these techniques in hybrid approaches or with context-aware information. The details of each paper are shown in the first column of Table \ref{generaltable2}, where we can observe that traditional techniques represent 48\%, 16\% and 13\% of the works, respectively for CF, CB and KB, hybrid approaches and context-aware RSs represent 25\% and 16\%, respectively.
	
Regarding the content-based technique, \citep{kofler2011} developed a prototype called \enquote{Near2me} integrating multimedia content items, user-generated metadata as their context to convey authenticity, and personalisation to the user.
	
We can highlight some projects that worked with collaborative filtering methods, like \citep{petrevska2012}, which developed the national tourism web portal in Macedonia, adopting the cloud-model CF to reduce the dimensionality of data and avoid the strict matching of attributes in similarity computation. \citep{jiang2011} presented a method named ContextRank, that calculates personalised interests for a specific user from different aspects, namely visual similarity score, textual tags similarity score and collaborative filtering score, which exploits different context information of geotagged web photos to perform personalised tourism recommendation. \citep{chang2013} calculated the similarity among users and users' network, combining collaborative filtering techniques, based on users appraisal and trustability evaluations, and social recommendations based on users' activities on SNs. Finally, other relevant models for collaborative filtering include the use of data mining models such as clustering, classification or association pattern mining, like in \citep{sun2013} and \citep{cao2010}.
	
The Knowledge-Based technique was used in three projects. For example, \citep{lucchese2012} proposed an algorithm for the interactive generation of personalised recommendations of POIs based on the knowledge mined from Flickr photos and Wikipedia. \citep{baraglia2012} created, on one hand, a knowledge model, used for calculating suggestions, and used, on the other hand, information of the path of a current user during a visit, that combined with the first one, allowed the system to produce a list of suggestions as possible locations to visit.
	
In addition to these traditional techniques, we highlight hybrid RSs. For instance, \citep{shen2015} used techniques such as content-based, semantic-based and social-based knowledge; \citep{meehan2013} in their hybrid project use collaborative filtering, content-based recommendation and demographic profiling. \citep{kurashima2010} introduced a hybrid RSs combining the Markov Model (using a probabilistic model that can handle sequential information) and topic models, also known as a hierarchical probabilistic model, in which a user is modelled as a mixture of topics, and a topic is modelled as a probabilistic distribution over landmarks.
	
Regarding context-aware systems, we can mention \citep{vakeel2016} who proposed an algorithm approach that applies a post-filtering contextual approach on a list of recommendations generated by traditional RS algorithms. Also, \citep{smirnov2013} developed TAIS, a mobile application that used an attraction information service, a recommendation service, a region context service, a ride-sharing service, and a public transport service. Another interesting work is the SPETA project \citep{garcia-crespo2009} that makes use of a variety of techniques which include context-aware, knowledge-based and social-based methods to retrieve the most suitable services. Finally, we highlight \citep{canneyt2011} who explored the possibility of using temporal context factors to better predict which POIs might be interesting to a given user.

\subsection{What properties of recommender systems are used?}
We surveyed a range of properties that are commonly considered when deciding which recommendation approach to select. As different applications have different needs, it must be decided which properties are important to pursue the specific application at hand. In this survey, we have identified the following properties: accuracy, coverage, confidence, trust, novelty, serendipity, diversity, utility, robustness and scalability, as defined by \citep{ricci2011} and shown in the third column of Table \ref{generaltable2}.

As expected, accuracy, which is one of the most fundamental measures through which RSs are evaluated. The main components of accuracy evaluation are: designing the accuracy evaluation; and accuracy metrics (accuracy of estimating ratings and accuracy of estimating rankings). In summary, accuracy is able to tell if the RS is able to predict those items that you have already rated or interacted with, thus RSs which optimize accuracy will naturally place those items at the top of a user's list, is found in almost all the projects analysed (77\% of them). The second most-seeked property is confidence, that can stem from available numerical values that describe the frequency of actions, i.e. how much time the user watched a certain show or how frequently a user bought a certain item. These numerical values indicate the confidence in each observation. Various factors that have nothing to do with user preferences might cause a one-time event; however, a recurring event is more likely to reflect user opinion \citep{herlocker2004}. That is, a confidence measure is important as it can help users decide which movies to watch, products to buy, and also help an e-commerce site in making a decision on which recommendations should not be displayed, because an erratic recommendation can diminish the trust of users in the system \citep{koren2009}.
	
In contrast, some projects concentrated in developing a recommender with the focus on a less \enquote{popular} property, such as \citep{memon2015}, oriented in improving scalability, that can be understood as the ability of the system to process an increasing amount of work with respect to a desirable performance metric, for example the predictive accuracy of the system \citep{barbosa2014a}. The importance of scalability has become particularly great in recent years because of the increasing importance of the \enquote{big-data} paradigm. A variety of measures are used for determining the scalability of a system: training time (Most RSs require a training phase, which is separate from the testing phase), prediction time (Once a model has been trained, it is used to determine the top recommendations for a particular customer), memory requirements (When the rating matrices are large, it is sometimes a challenge to hold the entire matrix in the main memory) \citep{ricci2011}. Or \citep{smirnov2013}, centred in guaranteeing robustness that means, an RS is stable and robust when the recommendations are not significantly affected in the presence of attacks such as fake ratings or when the patterns in the data evolve significantly over time. In general, significant profit-driven motivations exist for some users to enter fake ratings, for instance, the author or publisher of a book might enter fake positive ratings about a book at Amazon.com, or they might enter fake negative ratings about the books of a rival.
	
In many cases, several properties are pursued. For instance, \citep{lemos2012} tried to improve both accuracy and confidence, to make satisfactory recommendations of georeferenced photos without prior knowledge of the user profile, considering only its current context; Also to analyse, the context in which the photos were taken is relevant in making recommendations; and the usage of a context model considering various contextual dimensions may lead to an improved recommendation comparing to the result of one which uses only one context attribute (e.g., location). Others combined accuracy with coverage \citep{baraglia2012,popescu2011,choudhury2010}, that is, even when an RSs is highly accurate, it may often not be able to ever recommend a certain proportion of the items, or it may not be able to ever recommend to a certain proportion of the users (this measure is referred to as coverage). Due to this limitation the trade-off between accuracy and coverage always needs to be incorporated into the evaluation process. There are two types of coverage, which are referred to as user-space coverage and item-space coverage, respectively.
	
Some of the properties can be traded-off, for instance, perhaps the decline in accuracy may imply that other properties (e.g. diversity) are improved. Besides, while we can certainly speculate that users would like diverse recommendations or reported confidence bounds, it is essential to show that this property important in practice. In other words, when suggesting a method that improves one of this properties, one should also evaluate how changes in this property affects the user experience, either through a user study or through online experimentation \citep{shani2011}.
	
Overall, independently of the property (or properties) seeked in the several RS that we have reviewed, it is clear that the diversity and quantity of the properties used in the scientific researches is increasing, which demonstrates that those features can improve even more the recommenders, when they are well applied.

\subsection{Which type of recommendation is generated?}
In this survey, we have found that RSs mainly generate three types of outputs: places/points of interest (\textit{POIs}) such as monuments, churches, museums, etc.; tourist routes inside or outside the cities (\textit{route}); and basic information or instructions of a tour, mountain walks, schedules (\textit{guide}). The output of each analysed project can be observed in the fifth column of Table \ref{generaltable2}.

Firstly, the most common output are \textit{POIs}, which represent 61\% of the projects analysed. \citep{jiang2011} for instance, proposed a new method called ContextRank, which exploits different context information of photos to recommend personalised tourism POIs. Their architecture first detects landmarks from geotagged photos and estimates their popularity; then, by analysing the photos and their textual tags, only the representative ones are extracted for each landmark. It calculates user similarity from users' travel histories with all this contextual information, predicts a user's preference score in a landmark from different aspects, and combines these scores to give the final recommendation of POIs with their proposed algorithm, called ContexRank. Another example of POIs recommendation is \citep{cao2010}, whose project generates recommendations based on visual matching and minimal user input, by creating clusters of geotagged images and then recommending those POIs matching a query input by the user describing her preferred destinations. Another one is presented by \citep{oku2014}, that proposed a method for mapping geotagged tweets to POIs on the basis of the substantial activity regions of the POIs as learned using one-class support vector machine. We also highlight \citep{phan2014}, that applies collaborative recommendation algorithms to geotagged photos in order to produce personalised suggestions for POIs in the geocoordinate space. They used a collection of 3 million Flickr geotagged photos on which a series of steps was applied: first, unique locations were identified by discretizing the continuous geocoordinates into geographic virtual bins; second, implicit feedback was calculated in a user/location matrix using normalised frequency; and third, missing feedback values were imputed through four different algorithms.

Secondly, we find that 26\% of projects recommend \textit{routes}, among which we highlight three works. The first one is presented by \citep{sun2013}, who developed a travel recommendation approach integrating landmark and routing. The routing is generated based on the Dijkstra algorithm, combined with spatial clustering of images. The second one is presented by \citep{okuyama2013}, which proposes a travel route RS based on sequences of geotagged photos. The authors explain that the online processing of the system consists of the following steps: selection of tourist places that a user would like to visit; presentation of travel route candidates; and presentation of the selected travel route on a map. The third one, unlike the two projects previously mentioned, \citep{yin2012}, developed not a recommendation of routes, but of pedestrian tracks of paths (remind that a path can be, for instance, \enquote{pedestrian path} in open areas without pre-established paths, such as a large garden), in this case for the \textit{Forbidden City} in China, helping users to plan trips. As an output, their recommender also shows some features like the distribution of the visit duration along with the path. Another feature is the popularity of a destination by the total number of paths of the destination; with this popularity, the system can recommend what the hottest destinations are, in terms of seasons or months, thus being able to tell users whether March or October is the best travel time, for instance.

Finally, we found systems recommending \textit{guide} in 13\% of projects. An example was the project developed by \citep{popescu2011}; according to the authors, classical tourist guides are usually organised around landmark popularity and fail to account for each visitor's preferences. Considering this issue, this project introduced techniques like collaborative filtering for personalising the visit guides, based on one's tagging record and on the discovery of users with similar preferences.	
	
\subsection{Which evaluation methods are used?}
In this section we classify the projects analysed in two possible evaluation methods, online or offline, presented by \citep{beel2013, beel2015, mandel2016}. On the one hand, the \textit{online} evaluations, recommendations are shown to real users of the system during their section, that is, the process of evaluating a system is generated with the active and direct participation of the users, where the investigator obtains real feedback from them. On the other hand, \textit{offline} evaluations use pre-compiled offline datasets from which some information has been removed, in other words, the process of evaluating a system is not developed with the active and direct participation of users, but rather, they can use data from users (real data), or not (synthetic data). Subsequently, the recommender algorithms are analysed on their ability to recommend the missing information \citep{beel2013}.

According to \citep{weigl2011}, although the number of studies that use users has increased, the conducting such studies on real-world remains time-consuming and expensive, particularly for academic researchers. Consequently, relatively few studies measuring aspects related to user satisfaction have been published \citep{ricci2011}.
	
In one hand, from all the papers analysed in this research (column 3 in Table \ref{generaltable2}), 84\% have evaluated their systems using the \textit{offline} method, such as \citep{memon2015}. They used a sample of a dataset from Flickr with 1,376,886 photographs with their spatial and temporal context, and cleaned these photos' data, removing two types of photos from dataset: photos that were collected in the result of search based on text containing name of a city in their metadata and photos with incorrect temporal context. Then, they applied the density-based clustering algorithm to geo-tags associated with photos. This way, they compared some methods like, popularity rank, collaborative filtering rank, classic rank and, recommend popular places, to show the effectiveness of context ranking, which is his propose. With this method of evaluation, Memon demonstrated that his project is able to predict tourist's preferences in a new city more precisely and generate better recommendations as compared to other recommendation methods.
	
Using no one, but three different datasets, tourist spots (Foursquare), geotagged tweets (Twitter), and geotagged photographs (Panoramio), \citep{oku2014} conducted qualitative analyses in order to evaluate the effectiveness of the proposed methods (mapping geotagged tweets to tourist spot and extracts features of the tourist spot). Thus, he showed the effectiveness of the methods through qualitative analyses.
	
Another example was proposed by \citep{jiang2011}, a method named ContextRank that used a dataset from Panoramio, containing approximately 15 million of geotagged photos. For each landmark, he choose 10 representative photos by clustering. In his offline evaluation, he compared his method to scale space representation of all the geotags proposed in \citep{clements2010}. His results showed that different kinds of context information can help to enhance the recommendation performance when a user is lack of travel history.
	
Unlike previous works, \citep{baraglia2012}, to build a knowledge model, chose to measure the effectiveness and the efficiency of the proposed solution using two trajectory sets: synthetic and real data. In relation to the offline real dataset, it was made up of data coming from Flickr, where the trajectories are built using users' photos. On the other hand, the offline synthetic dataset was generated using a trajectory generator for a specific geographic area. It takes as input a dataset of POIs, which are combined in sequences that form trajectories. In this way, this project was able to perform two evaluations: (1) the quality of the trajectory set, adopting spatial coverage, data coverage, region separation and rate; and (2) the effectiveness and efficiency adopting the prediction rate, accuracy, average error and omega. The results showed that this project is able to generate suggestions of potential POIs, depending on the current position of a tourist, and a set of trajectories describing the paths previously made by other tourists.
	
On the other hand, only 16\% of the projects have submitted their projects to an assessment by real users. Even those counted on a very low amount of users. Using 21 participants represented in 8 different countries, \citep{tintarev2010} developed an online evaluation to see the effect of personalisation on the behaviour of participants. Besides to use a questionnaire to get travel habits of the participants who started travelling in the past, they had also completed parallel data collection tasks. Then, a list of recommendations of POIs was generated. These lists were either personalised or based on popularity, but both consisted of precisely five POIs given the limited time available for sightseeing. These participants received a list of recommended points of interest to say how much they liked of each POI on a scale from 1 to 7 (1=not at all, 7=a lot). Although some participants did not follow many of the recommendations in the personalised lists, the author found that personalised recommendations enabled a \enquote{discovery mode}, that is, participants visited more POIs than in the popular condition, and these POIs were also rarer than POIs visited by participants in the popular condition. Thus, this project showed that personalised recommendations may increase serendipity since users are more likely to discover sites that surpass their a priori assessment.
	
In the project called MMedia2U developed by \citep{lemos2012}, a group of 13 users evaluated photos from 8 different contexts, each one consisting of a stage of evaluation. Lemos pointed out in his project that an online evaluation of an RS is a hard task, due to the fact that item's relevancy has a strongly personal nature and it is complex to be measured. This difficulty is enhanced when existing a lack of historical evaluation data, which makes large-scale studies very costly and difficult to be run. In his case, the complexity is even bigger, since his project needs to range the possible contexts of real situations. In each stage of his evaluation, one context (approximately 100 photos, 20 were taken in similar contexts to the one showed to the user and 80 were different in some dimensions of the context) was presented to the user. The volunteers had to visualize a set of photos and choose those that seemed to be more appealing to him/her, taking into consideration the context he/she suggested. And then, the degree of success on recommendations was then evaluated by the ratio of chosen photos. In general, the results of this project concluded that, for the data used, context-awareness can bring gains in the photo recommendation compared to a random list.
	
In the case of \citep{kofler2011}, 12 volunteers participated in the user-oriented evaluation of the prototypical implementation of Near2me. This project focused in discovering: how Near2me is perceived by users in general and how users interact with the system; how are the individual components used to contribute to the users' satisfaction with the system; and finally, how the interplay of the components used convey authenticity and personalisation to the user. The evaluation consisted of a task-directed walkthrough of the interface carried out on the working prototype. During the evaluation, the subjects were asked to use the Near2me prototype to plan a possible trip to Paris and were left free to interact with the prototype for a maximum time of 30 minutes. While performing the task, the participants were asked to speak aloud, giving insights about the motivations behind each action, the possible expectations about the foreseen outputs, and the satisfaction towards the actual recommendation and interaction paradigm. The subjects were observed, most relevant comments and behaviours were noted, and each session was recorded using both a video camera and screencast software. After the walkthrough, information was obtained from the participants through semistructured interviews. A question framework based on the research questions guided the interviews. This framework was adapted for each participant according to her vocabulary and the notes were taken during observation allowed for exploring and confirming the participant's feedback. This evaluation showed that the participants are interested in three perspectives: locations, topics, and experts.

\subsection{What type of interface is used?}
In our survey, we have found projects that use an interface based on mobile phones, based on web or without any interface at all. Specifically, from 18 papers that provided an interface, 12 of them (67\%) were web-oriented and the remaining (33\%) were mobile-oriented. These are detailed in column 7 (Table \ref{generaltable2}). We did not find any desktop-oriented application.
	
\begin{figure}[h]
	\centering
	\includegraphics[scale=0.45] {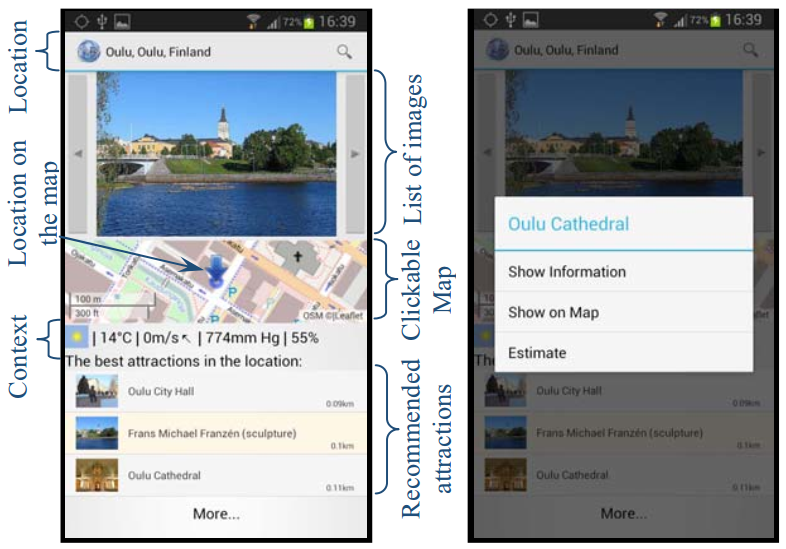}
	\caption{Mobile interface of TAIS: main screen, context menu with actions}
	\label{Interface_Smirnov2013}
\end{figure}

An example of an RS with an interface for Android mobile phones is the app TAIS (Tourist Assistant) developed by \citep{smirnov2013}. The main application screen is shown in Figure \ref{Interface_Smirnov2013} (left). The tourist can see images extracted from accessible internet sources, a clickable map with his/her location, current weather, and the attractions around ranked by the recommendation service. When the tourist clicks on an attraction, a context menu shows detailed information about the chosen attraction (Figure \ref{Interface_Smirnov2013} right).
	
\begin{figure}[h]
	\centering
	\includegraphics[scale=0.32] {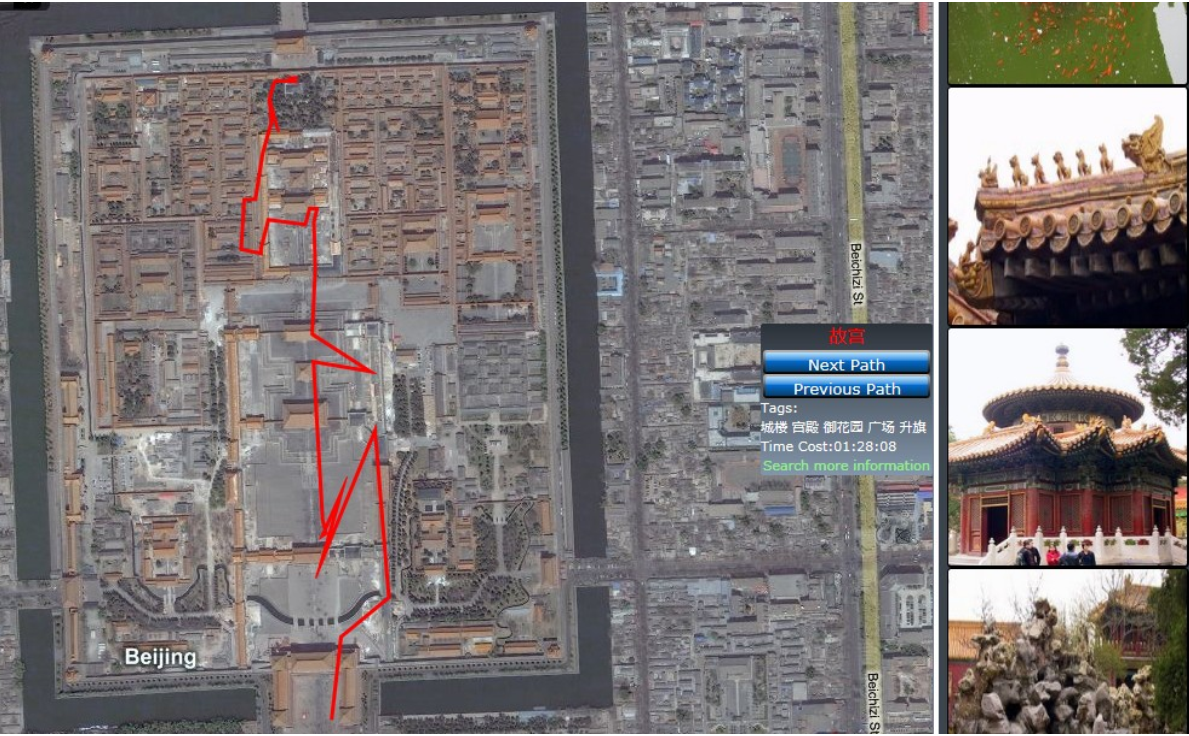}
	\caption{An interface of path recommendation of user-specified places.}
	\label{Interface_Yin2012}
\end{figure}

We also show some details of the web-oriented project presented by \citep{yin2012}, which, unlike the other projects, makes a recommendation of not only where to visit but also how to visit, that is, it makes a recommendation of \enquote{path} alongside with high-quality photos taken in this destination. In Figure \ref{Interface_Yin2012}, we see an example of the results obtained after a user inputs a destination name and then get the recommended paths within the query destination, in this case \enquote{the Forbidden City}.
	
\begin{figure}[h]
	\centering
	\includegraphics[scale=0.4] {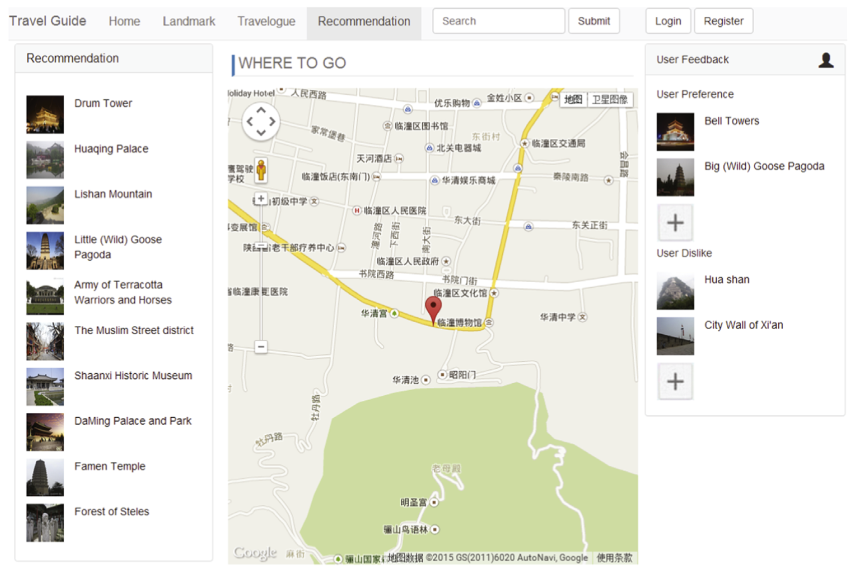}
	\caption{User's interface is shown a visual examples in Xi'an, China.}
	\label{Interface_Shen2015}
\end{figure}

A web-oriented project was introduced in \citep{shen2015}. Figure \ref{Interface_Shen2015} shows a visual example of the personalised travel recommendation. The system can collect the current location and show the located city on the map with high-quality photos taked in that destination are also shown to users. Also, the user can input their favourite and non-favourite attractions on the right side of the interface. If the user does not wish to interact with the system, the system will show them the results which are ranked by popularity, to avoid the cold-start problem.

\section{Discussion}
As we showed in this article, the combination of RSs and SNs is obtaining better results and, indirectly, enhancing the tourism sector's economy \citep{jia2016}. It is crucial, since the application of RSs in such a customer sensitive sector has become a necessity, not a luxury; moreover, RSs have great value because they assist all parts of the tourism value chain. On one side, they support better and faster decisions when the customer is choosing a destination, and help them to plan holidays according to their needs, improving the overall service offered. On the other side, they also offer considerable benefits for service providers such as hotels, restaurants or cultural event organisers, improving their online presence, increasing sales, and reducing costs for advertising activities \citep{rabanser2005}.

This way, the extent of projects that use SNs in their RSs keeps growing, as well as the volume of data generated in those environments, as shown in Fig. \ref{2020}, thus progressively influencing tourists around the world. The RSs are deeply changing the way tourists search, find, read and trust when choosing a destination. On the other hand, people through SNs create and share content related to everything, from travel agencies to relevant information about a certain POI. However, the increment in the academic research production can be affected by some relevant challenges, from which the main is to get access to the data from SNs.
	
In 2018, Facebook, for instance, announced dramatic data access restrictions on its app and website in response to the public outcry following the Cambridge Analytica scandal \citep{confessore2018}. This decision made it virtually impossible to carry out large-scale research on Facebook. The changes make extinct software and libraries dedicated to academic research on Facebook, including Netvizz, NodeXL, SocialMediaLab, fb\_scrape\_public and Rfacebook, all of which relied on Facebook's APIs to collect data. In the case of Twitter, it operates three well documented public APIs, in addition to its premium and enterprise offerings. Twitter's relative accessibility leads it to be vastly overrepresented in social media research. But public and open APIs are an exception in the social media ecosystem. Facebook's Public Feed API, for example, is restricted to a limited set of media publishers.

Due to the increasing data restriction on the part of the large companies such as Twitter \citep{cameron2015}, Instagram \citep{bright2017}, and Facebook \citep{seetharaman2015}, some campaigns and initiatives pro data sharing have gaining adepts in the scientific environment. The idea of one of those projects, known as \enquote{Open Data}\footnote{http://www.opendatafoundation.org/} is that the data be available for everyone, without restrictions, and can be freely used, reused and redistributed by anyone, meeting the requirement of mentioning the original source and sharing under the same licenses in which that information was collected. In other words, the goal of the open data movement is similar to others such as open source, open content and open access.

We believe that data sharing, whether from SNs, public or private bodies, is extremely relevant for the researchers in all areas of knowledge. In the case of public bodies, the Ministers of Science of all nations belonging to the Organisation for Economic Co-operation and Development (OECD\footnote{http://www.oecd.org}) signed a statement in 2004 saying that, basically, all archive data publicly funded must be accessible for the public. With respect to the data available online like in SNs, future researches would have to deal with an increasingly sensitive and troubling phenomenon, the privacy and the use of the data, among other reasons because they have stored very intimate data. Recently, we can observe two simultaneous scenarios: the SNs that provide APIs to the data access and analysis; and the SNs that suppress it, such as Facebook as we have already mentioned.

According to \citep{bastos2018}, the aforementioned data restriction, which causes a differentiation between public and \enquote{premium} versions, will widen the gap between industry researchers hired by SNs and researchers working outside of corporations. 

In spite of such restrictions, there are large databases available for research purposes, which could be used on projects that seek an offline evaluation to measure their accuracy, for instance. Some examples of those databases are Open Data, Stanford Large Network Dataset Collection, UCI Network Data Repository, Interesting Social Media Datasets, Network data, and Kevin Chai's.

Throughout this paper, we tried to disclose and clarify some theoretical and technical topics in the development of a recommender, by analysing the projects of recommendation systems since mid-2004. We presented a summary of the basic recommendation techniques, an overview of what SNs are about, their benefits, and their importance to the recommendations projects. Then, we ordered (by date of publication) the main works of the last 10 years about RSs in the tourism sector that make use of SNs and classify them into categories such as: SNs and online databases used, items extracted from these sites, evaluation techniques applied, general goals in evaluating, display and interface.

Overall, we observed that RSs are diversifying their data source, consequently adding more complexity in its ability to interpret and predict the customer interests. There are still many researches that use a single data source (e.g. Flickr), which retrieve data considered basic (e.g. age, gender, marital status, number of children, etc.), and seek only the accuracy improvement by means of basic techniques such as CB and CF to generate POIs recommendations. But then, recent investigations started to use more complex data (e.g. correlations between network contacts in a SN, behaviour, texts, photos, etc.) from multi data sources (e.g. Facebook + Wikipedia + TripAdvisor) and different properties (e.g. novelty, serendipity, diversity), increasing the variety of assessments, mainly thanks to machine learning.
	
We also consider that the use of SNs (also known social-based RSs) can indirectly solve or at least mitigate some well-known issues of recommenders, such as the problem of (1) the new user/item, known as the cold start problem; (2) sparsity or ratio diffusion; (3) compilation of demographic information; (4) Portfolio effect; (5) recommendations with excessive results; (6) serendipity \citep{aggarwal2016, ricci2015, tavakolifard2012},as well as, to improve the quality of recommendations in the tourism context \citep{law2011}.

The cold start problem (1) appears with new users/items, i.e., a system is not capable of recommending an item with an acceptable accuracy until the user has rated enough items. By using SNs, this problem can be mitigated, since it is possible to retrieve \enquote{likes}, comments, and reviews made by the user in one or more SN. Similarly, there is the new element problem, in which a new item is not recommended until a considerable number of users have rated it, so the probability of the system recommending such item is low. To get around this problem, first, the POIs ratings could be retrieved from different SNs such as Facebook, Flickr, TripAdvisor or Google Maps. Secondly, those POIs with no enough reviews or comments can be of interest for people who like exotic, isolated or less known places; thus, if the system is able to detect those profiles, it would be able to recommend them those places.
	
The sparsity or ratio diffusion problem (2) occurs when there are few or no user ratios that match each other, thus there would be few users to compare with or few similar elements to look for. This problem is commonly found in CB and CF RSs. In this context, SNs play a crucial role due to its large extent of user profiles available, which could minimise or even neutralise such problem.
	
The compilation of demographic information (3) refers to the lack of information related to where people reside or is currently located. Sometimes, a user can be reticent in providing information to a new system, whether due to their privacy concerning, whether due to lack of trust in the service. The use of data already shared on SNs, also used to retrieve such kind of information in a non-intrusive way, could solve this problem.

Portfolio effect (4) is regarding the recommendation of an item very similar to another item that the user already has in her history. In the case of tourism, an RS that previously knows the places the user has visited through information posted on their SN could then avoid recommending places of similar categories and locations.

The recommendations with poor or excessive results (5) can overwhelm the user. In order to reduce or specialise the items recommended, additional properties could be applied to the RSs, such as novelty, diversity, serendipity, utility, etc. A good number of those properties could be based on the personality predicted using data available on SNs, following some already existent psychological theories. For instance, the system could recommend useful POIs in a reduced quantity when considering the curiosity, that means, the higher the degree of curiosity, the lower the popularity of the POI, and vice-versa \citep{menk2017curumim2}.

One of the keys of the serendipity (6) may be the prediction of an individual's personality. By using data from SNs, such prediction could be easily achieved, thus the RSs would be able to positively surprise the user by recommending items that really match the user's interests. Regarding the adoption of real users to assess the recommenders developed, it is worthwhile stressing its importance when measuring the quality of a system. It is highly abstract to build a system that generates positive surprisingly (serendipitous) recommendations without the cooperation of a human being since each has unique tastes, and the same item may be relevant for one individual but not for another. In short, the researcher needs to understand the response of the user to the delimited parameter, which is not feasible in an offline environment.

In spite of the advantages and facilities that offline tests offer to the researcher, we believe that a recommendation system shall be submitted to field experimentation, where the data are recorded from reactions resulting of the variables the researcher enter in the experiment; as previously stated, the variables are not controlled, because the RSs are developed for human beings, whose tastes, situations and profiles are different. To strengthen this point of view, we must also analyse the psychological relationships between tourism and psychology \citep{jani2011} or recommendation projects that use psychology to improve their recommendations \citep{chen2013, roshchina2015, ortigosa2014}. We believe that the RSs cannot lose their target, which is the human being and the context in which it is presented. That is, the individual plays an extremely important role in this process.

Nevertheless, projects counting on the participation of volunteers to assess their systems, thus seeking an online evaluation, have as possibilities the Open Source Social Network (OSSN), a rapid development social networking software, but then it would be needed to recruit volunteers to feed those OSSNs, which is laborious. Another option for projects that need user interaction is Diaspora\footnote{https://diasporafoundation.org/}, an SN launched in 2010 that already has 600 thousand users, where the user \enquote{owns} his data and has the power to share it as he wants. Therefore, with the request and acceptance, these data could be used.
	
Although further studies are needed to assess the benefits of the online evaluation, it is vital to encourage the forthcoming projects to ask for feedback from the users, who are the main beneficiaries of the recommenders. This way, it would be possible to widely explore the influence and the impact of SNs in all the aspects of the RSs in the tourism sector.

We expect that the clarification of which SNs were used in the recommendation projects may contribute in encouraging the use of SNs as a method of nourishing their RSs in new projects, since nowadays its use is simple and accessible to any researcher. In general words, we hope to contribute to make an approach about the recommendation systems and SNs to cover existing definition in the literature, their types and characteristics; we also hope that the state-of-the-art knowledge here generated can support researchers and practical professionals in their understanding of developments in RS applications.


With regard to the challenges of future investigations, it is important to emphasise that we did not find works of RSs for the sector of tourism that use human personality to enrich the user profile so that different aspects can be taken into account \citep{bologna2013, mairesse2007}. Also, we consider that the generation of recommendation in the tourism sector based on SNs and that somehow consider the human personality will have a start of importance. In this sense, the first steps have already been taken in other areas of knowledge in the industry, and this will not be different.

\begin{acknowledgements}
	This work was supported by CAPES Foundation, Ministry of Education of Brazil, Brasilia - DF, Zip Code 70.040-020 (Alan Menk) and Spanish Government Project TIN2017-88476-C2-1-R (Laura Sebastia).
\end{acknowledgements}
		
\nocite{yin2012}
\nocite{okuyama2013}
\nocite{mahmood2013}
\nocite{canneyt2011}
\nocite{kofler2011}
\nocite{choudhury2010}
\nocite{garcia-crespo2009}
\bibliographystyle{spbasic} 
\bibliography{biblio.bib} 
\end{document}